\documentclass[
    ,final            
  ]
  {aipproc}

\layoutstyle{6x9}

\usepackage{graphics}
\usepackage{psfrag}
\usepackage{epsfig}
\usepackage{psfig}
\usepackage{epsf}
\usepackage{float}

\newcommand{\beq}{\begin{equation}}
\newcommand{\eeq}{\end{equation}}
\newcommand{\beqa}{\begin{eqnarray}}
\newcommand{\eeqa}{\end{eqnarray}}
\newcommand{\nn}{\nonumber \\ }
\newcommand{\fet}[1]{\mbox{\boldmath $#1$}}

\begin{document}

\title{Effective Field Theory and Isospin Violation in Few--Nucleon Systems}

\author{E.~Epelbaum}{
  address={Jefferson Laboratory, Theory Division, Newport News, VA 23606, USA}
}

\begin{abstract}
I discuss the leading and subleading isospin--breaking three--nucleon forces 
in the chiral effective field theory framework.

\end{abstract}

\maketitle


\section{Introduction}

Isospin violation has its origin within the Standard Model in 
the different masses of the up and down quarks and electromagnetic interaction.
Its consequences for few--nucleon systems can be studied in a systematic way within 
chiral Effective Field Theory (EFT). This approach is based on the most general (approximately) chiral invariant 
Lagrangian for pions and nucleons which 
includes all possible interactions consistent with the isospin violation in the Standard Model.
In particular, strong isospin--breaking terms are proportional to $\epsilon  M_\pi^2$, where 
$\epsilon = ( m_d-m_u )/ (m_d+m_u ) \sim 1/3$. Electromagnetic vertices due to exchange of (hard) virtual photons 
are proportional to the nucleon charge matrix $Q_{\rm ch}= e/2 \, (1 + \tau_3 )$. 
In principle, one should also include explicit soft photons. Their contributions are, however,
irrelevant for the present study and will not be considered.
The effective Lagrangian has been applied to study isospin--violating  two--nucleon (2N) forces 
\cite{VanKolck:1993ee,vanKolck:1996rm,vanKolck:1997fu,Epelbaum:1999zn,
Walzl:2000cx,Friar:1999zr,Friar:2003yv,Friar:2004ca}.
In these proceedings, I consider isospin--breaking three--nucleon forces (3NFs) within the chiral EFT approach, 
see also \cite{Epelbaum:2004xf,Friar:2004rg}.

\section{Power counting and the effective Lagrangian}
\label{sec:pc}

In this work, I use the same counting rules for $e$ and $\epsilon$ as in \cite{Epelbaum:2004xf}, namely:
\beq\label{CountRules}
\epsilon \sim e \sim \frac{q}{\Lambda}; \quad \quad
\frac{e^2}{(4 \pi )^2}  \sim \frac{q^4}{\Lambda^4}\,,
\eeq
where $q$  ($\Lambda$) refers to a generic low--momentum scale (the pertinent hard scale). 
The N--nucleon force receives contributions of the order $\sim (q/\Lambda )^\nu$,
where 
\beq
\label{powc}
\nu = -2 + 2 N -2 C + 2 L + \sum_i V_i \Delta_i\,.
\eeq
Here, $L$, $C$ and  $V_i$ refer to the number of loops, separately connected pieces and vertices of type $i$,
respectively. Further, the vertex dimension  $\Delta_i$ is given by
\begin{equation}
\label{chirdim}
\Delta_i = d_i + \frac{1}{2} n_i - 2\;,
\end{equation}
where  $n_i$ is the number of nucleon field operators and $d_i$ is the $q$--power of the 
vertex, which accounts for the number of derivatives and insertions of pion mass, $\epsilon$ and
$e/(4 \pi )$ according to eqs.~(\ref{CountRules}). Notice further that the nucleon mass is counted 
according to $q/m \sim (q/\Lambda )^2$, see \cite{Epelbaum:2004xf} for more details.

The relevant isospin--symmetric terms in the effective Lagrangian in the nucleon rest frame are 
\cite{Bernard:1995dp,Fettes:2001cr}:
\beqa
\label{lagr}
\mathcal{L}^{(0)} &=& \frac{1}{2} \partial_\mu \fet \pi \cdot \partial^\mu \fet \pi  - \frac{1}{2} M_{\pi}^2 \fet \pi^2  
+ N^\dagger  \Big(  i \partial_0 +
\frac{g_A}{2 F_\pi} \fet \tau \vec \sigma \cdot \vec \nabla \fet \pi 
- \frac{1}{4 F_\pi^2} \fet \tau \cdot ( \fet \pi \times \dot{\fet \pi } ) \Big) N \nn
&& {}- \frac{1}{2} C_S ( N^\dagger  N  )  ( N^\dagger  N  )  - \frac{1}{2} C_T  
( N^\dagger \vec \sigma N )  ( N^\dagger \vec \sigma N ) \nn
\mathcal{L}^{(1)} &=& N^\dagger  \Big(  
- \frac{2 c_1}{F_\pi^2} M_\pi^2 \fet \pi^2 + \frac{c_3}{F_\pi^2} (\partial_\mu \fet \pi \cdot \partial^\mu \fet \pi ) - 
 \frac{c_4}{2 F_\pi^2} \epsilon_{ijk} \, \epsilon_{abc} \, \sigma_i \tau_a (\nabla_j \, \pi_b ) (\nabla_k \, \pi_c )  
 \Big) N \,, \nn
&& {}- \frac{D}{4 F_\pi}   ( N^\dagger N  )  ( N^\dagger  \vec \sigma \fet \tau  N  ) \cdot \vec \nabla \fet \pi
\eeqa
where $M_{\pi}$ and $F_\pi$ refer to the pion mass and decay constant, $g_A$ denotes the nucleon axial coupling 
and $c_{i}$, $C_{S,T}$ and $D$ are further low--energy constants (LECs). 
The relevant isospin--violating part of the Lagrangian reads \cite{Meissner:1997ii}:
\beqa
\mathcal{L}^{(2)} &=&  \frac{1}{2} \delta M_{\pi}^2 \pi_3^2  + N^\dagger \Big(- \frac{1}{2} \tau_3 \delta m   
-\frac{c_5}{F_\pi^2} \epsilon M_\pi^2 (\fet \pi \cdot \fet \tau ) \pi_3 
\Big) N \nn
\mathcal{L}^{(3)} &=& N^\dagger  \Big( f_1 e^2 (\pi_3^2 - \fet \pi^2 ) 
+ \frac{1}{4} f_2 \, e^2 ( (\fet \pi \cdot \fet \tau ) \pi_3 - \fet \pi^2 \tau_3 ) \nn
&& \quad \quad  + \frac{2 d_{17} - d_{18} - 2 d_{19}}{F_\pi} \epsilon 
M_\pi^2 \vec \sigma \cdot \vec \nabla \pi_3 \Big) N 
+ L \epsilon M_\pi^2 ( N^\dagger \tau_3 N ) (N^\dagger N )\,,
\eeqa
$c_5$, $d_i$, $f_{1,2}$ and $L$ are the LECs and $\delta M_\pi^2 = M_{\pi^\pm}^2 - M_{\pi^0}^2$.
The LECs $c_5$ and $f_2$ are related to the proton--to--neutron mass difference 
via $(\delta m )^{\rm str.} \equiv (m_p - m_n )^{\rm str.} = -4 c_5 \epsilon M_\pi^2$ 
and  $(\delta m )^{\rm em.} \equiv (m_p - m_n )^{\rm em.} = - f_2 e^2 F_\pi^2$.
Further, $\delta m =(\delta m )^{\rm str.} + (\delta m )^{\rm em.}$.  Notice that 
terms in the last line of the above equation lead to vanishing contributions to the 3NF and
were not considered in \cite{Epelbaum:2004xf}. Here I decided to keep them for the sake of completeness. 

\section{Isospin--breaking three--nucleon force }
\label{sec:3nf}

\begin{figure}[htb]
\vspace{0.5cm}
\psfrag{x11}{\raisebox{-0.1cm}{\hskip -0.1 true cm  (a)}}
\psfrag{x22}{\raisebox{-0.1cm}{\hskip -0.1 true cm  (b)}}
\psfrag{x33}{\raisebox{-0.1cm}{\hskip -0.1 true cm  (c)}}
\psfrag{x44}{\raisebox{-0.1cm}{\hskip -0.1 true cm  (d)}}
\psfrag{x55}{\raisebox{-0.1cm}{\hskip -0.1 true cm  (e)}}
\psfrag{x66}{\raisebox{-0.1cm}{\hskip -0.1 true cm  (f)}}
\psfrag{x77}{\raisebox{-0.1cm}{\hskip -0.1 true cm  (g)}}
\psfig{file=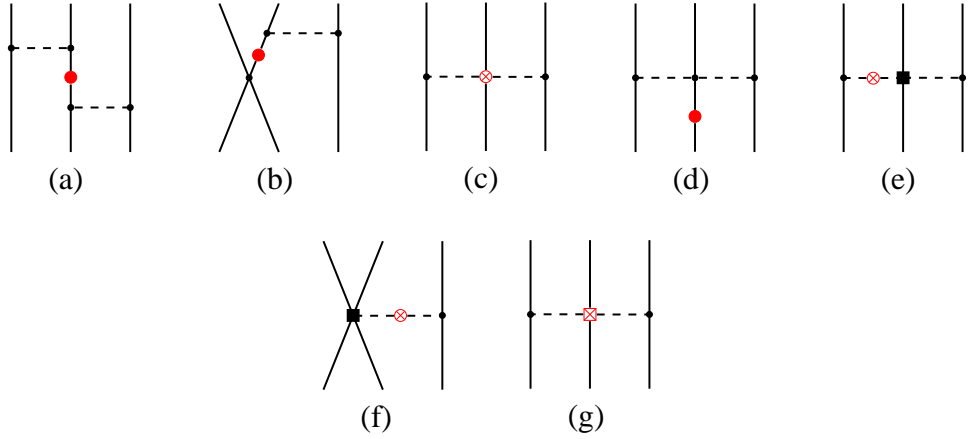,width=13.0cm}
\vspace{0.3cm}
\caption{\label{fig1} Leading (a--d) and subleading (e--g) isospin--violating contribution to the 3NF. 
Solid dots (filled rectangles) denote isospin--invariant vertices with $\Delta_i = 0$ ($\Delta_i = 1$) while
crossed circles (rectangles) refer to isospin--breaking vertices with $\Delta_i = 2$ ($\Delta_i = 3$).   
Filled circles refer to incertions of $\delta m$. Only one representative diagram of each kind is shown.
}
\vspace{0.5cm}
\end{figure}

The diagrams contributing to the leading ($\nu = 4$) and subleading ($\nu = 5$) isospin--breaking 3NFs 
are depicted in Fig.~\ref{fig1}. It should be understood that these diagrams only specify the topology
and do not correspond to Feynman graphs. Clearly, the contributions to the 3NF do not include the 
pieces generated by the iteration of the 2N potential. 
In \cite{Epelbaum:2004xf} we have evaluated the corresponding 3NFs using the method 
of unitary transformation developed in \cite{Epelbaum:1998ka}.  
For example, to calculate the contribution of the graph (a) in Fig.~\ref{fig1} one needs to evaluate the 3N matrix 
elements of the operator
\beqa
\label{operTPE}
V_{2 \pi} &=& \eta ' \bigg[ \frac{1}{2} H_{1} \frac{\lambda^1}{(H_0 - E_{\eta '})}  H_{1} \, \tilde \eta 
\,  H_{1} \frac{\lambda^1}{(H_0 - E_{\tilde \eta} )( H_0 - E_{\eta '} )}  H_{1} \nn 
&& \mbox{\hskip 0.7 true cm} -\frac{1}{8} H_{1} \frac{\lambda^1}{(H_0 - E_{\eta '})}  H_{1} \, \tilde \eta 
\,  H_{1} \frac{\lambda^1}{(H_0 - E_{\tilde \eta} )( H_0 - E_{\eta} )}  H_{1} \nn
&&  \mbox{\hskip 0.7 true cm} + \frac{1}{8} H_{1} \frac{\lambda^1}{(H_0 - E_{\eta '}) ( H_0 - E_{\tilde \eta} )}  
H_{1} \, \tilde \eta 
\,  H_{1} \frac{\lambda^1}{(H_0 - E_{\tilde \eta} )}  H_{1} \nn
&&  \mbox{\hskip 0.7 true cm} - 
\frac{1}{2} H_{1} \frac{\lambda^1}{(H_0 - E_{\eta})}  H_{1} \,  \frac{\lambda^2}{(H_0 - E_{\eta})} 
\,  H_{1} \frac{\lambda^1}{(H_0 - E_{\eta} )}  H_{1}
\bigg] \eta  + \mbox{h.~c.}\,,
\eeqa
where $\eta$, $\eta '$ and $\tilde \eta$ denote the projectors on the purely nucleonic subspace of the Fock space, while 
$\lambda^i$ refers to the projector on the states with $i$ pions. 
$E_\eta$, $E_{\eta '}$ and $E_{\tilde \eta}$ refer to the energy of the nucleons in the states $\eta$, $\eta '$ and 
$\tilde \eta$, respectively. 
Further, $H_1$ is the leading 
$\pi NN$  vertex $\propto g_A$ in eq.~(\ref{lagr}), and 
$H_0$ denotes the free Hamilton operator for pions and nucleons corresponding to the density
\beq
\label{freeH}
\mathcal{H}_0 = \frac{1}{2} \dot{\fet \pi} ^2 + 
\frac{1}{2} (\vec \nabla \fet \pi )^2 + \frac{1}{2} M_\pi^2  \fet \pi^2 +
\frac{1}{2} N^\dagger \delta m \tau_3 N\,.
\eeq
One finds the following charge--symmetry--breaking (CSB) 3NFs resulting from diagrams (a) and (b):
\beqa
\label{3NFisosp1}
V^{\rm 3N}_{2\pi} &=& \sum_{i \not= j \not= k} \,2 \delta m  \,  \left(
  \frac{g_A}{2 F_\pi} \right)^4 \frac{( \vec \sigma_i \cdot \vec q_{i}  ) 
(\vec \sigma_j \cdot \vec q_j  )}{(\vec q_i{} ^2 + M_{\pi}^2 )^2 ( \vec
q_j{} ^2 + M_{\pi}^2)} \bigg\{ [\vec q_i \times \vec q_j ] \cdot \vec \sigma_k  \, [ \fet \tau_i \times \fet \tau_j ]^3
 \nn
&& {} + \vec q_i \cdot \vec q_j \left[ (\fet \tau_i \cdot \fet \tau_k ) \tau_j^3  -
(\fet \tau_i \cdot \fet \tau_j ) \tau_k^3 \right] \bigg\}\, \nn
V^{\rm 3N}_{1\pi} &=& \sum_{i \not= j \not= k} 2 \, \delta m \, C_T   \left( \frac{g_A}{2 F_\pi} \right)^2
\frac{\vec \sigma_i \cdot \vec q_i}{(\vec q_i {} ^2 + M_{\pi}^2)^2} \, [ \fet \tau_k \times \fet \tau_i ]^3 \; 
[ \vec \sigma_j \times \vec \sigma_k ] \cdot \vec q_i \,,
\eeqa
where $i$, $j$ and $k$ denote the nucleon labels. Further, $\vec q_i \equiv \vec p_i \, ' - \vec p_i$
where  $\vec p_i$ ($\vec p_i \, '$) are initial (final) momenta of the nucleon $i$.
The  $2\pi$--exchange diagrams (c), (d) and (g) in Fig.\ref{fig1} lead to the 3NF
\beqa
\label{3NFprom}
V_{2 \pi}^{\rm 3N} &=& \sum_{i \not= j \not= k}  \bigg(
  \frac{g_A}{2 F_\pi} \bigg)^2 \, \frac{( \vec \sigma_i \cdot \vec q_{i} ) 
(\vec \sigma_j   \cdot \vec q_j  )}{(\vec q_i{}^2 + M_{\pi}^2 ) ( \vec
q_j{}^2 + M_{\pi}^2)} \nn
&& \mbox{\hskip 2 true cm} \times
\bigg[ \frac{( \delta m )^{\rm str.}}{4 F_\pi^2} \Big( 2 (\fet \tau_i \cdot \fet \tau_k ) \tau_j^3 - 
(\fet \tau_i \cdot \fet \tau_j ) \tau_k^3  \Big) + f_1 e^2 \tau_i^3 \tau_j^3 \bigg]\,.
\eeqa
Finally, the charge--symmetry--conserving 3NF resulting from diagrams (e) and (f) in Fig.~\ref{fig1} reads:
 \beqa
\label{3NFisosp4}
V^{\rm 3N}_{2\pi}&=&\sum_{i \not= j \not= k} \, \delta M_\pi^2 \, \left(
  \frac{g_A}{2 F_\pi} \right)^2 \frac{( \vec \sigma_i \cdot \vec q_{i})
(\vec \sigma_j   \cdot \vec q_j  )}{(\vec q_i{}^2 + M_{\pi}^2 )^2 ( \vec
q_j{}^2 + M_{\pi}^2)}  \bigg\{ \tau_i^3 \tau_j^3 \left[  - 
\frac{4 c_1 M_\pi^2}{F_\pi^2} +    \frac{2 c_3}{F_\pi^2} (\vec q_i \cdot \vec q_j )   \right] \nn
&& \mbox{\hskip 1.7 true cm} 
+ \frac{c_4}{F_\pi^2} \, \tau_i^3 \, 
[\fet \tau_j \times \fet \tau_k ]^3 \, [\vec q_i \times \vec q_j ] \cdot \vec \sigma_k \bigg\} \nn
V^{\rm 3N}_{1\pi}&=&- \sum_{i \not= j \not= k} \, \delta M_\pi^2 \, \frac{g_A}{8 F_\pi^2} \, D \, 
\frac{\vec \sigma_i \cdot \vec q_i }{(\vec q_i{}^2  + M_\pi^2 )^2}  
\,  \tau_i^3 \tau_j^3 (\vec \sigma_j \cdot \vec q_i )\,.
\eeqa
These expressions should be used together with the corresponding isospin--symmetric 3NFs expressed in terms 
of the charged pion mass.

\begin{figure}[htb]
\psfrag{x11}{\raisebox{-0.1cm}{\hskip 0.0 true cm  (a)}}
\psfrag{x22}{\raisebox{-0.1cm}{\hskip 0.0 true cm  (b)}}
\psfrag{x33}{\raisebox{-0.1cm}{\hskip 0.0 true cm  (c)}}
\psfrag{x44}{\raisebox{-0.1cm}{\hskip 0.0 true cm  (d)}}
\psfrag{x55}{\raisebox{-0.1cm}{\hskip 0.0 true cm  (e)}}
\psfrag{x66}{\raisebox{-0.1cm}{\hskip 0.0 true cm  (f)}}
\psfrag{x77}{\raisebox{-0.1cm}{\hskip 0.0 true cm  (g)}}
\psfrag{x88}{\raisebox{-0.1cm}{\hskip 0.0 true cm  (h)}}
\psfrag{x99}{\raisebox{-0.1cm}{\hskip 0.0 true cm  (i)}}
\psfig{file=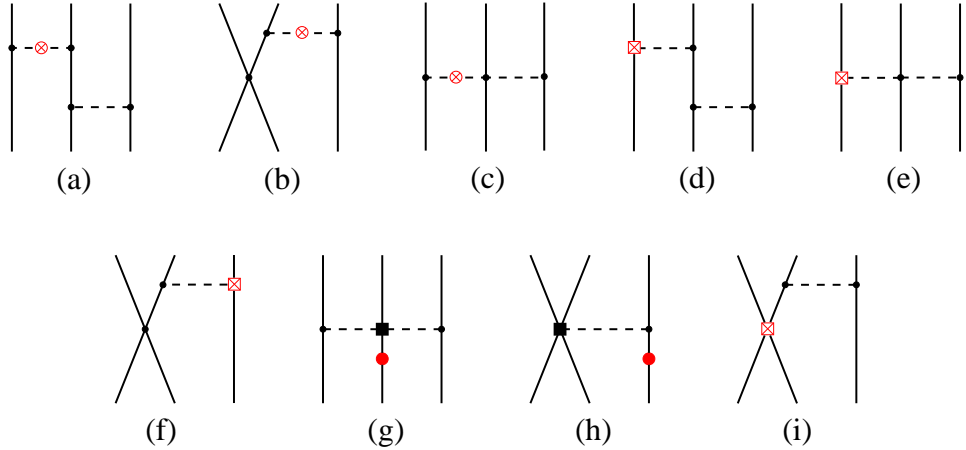,width=13.0cm}
\vspace{0.3cm}
\caption{\label{fig2} Leading (a--c) and subleading (d--i) isospin--violating contribution to the 3NF which vanish,
as discussed in the text. For notation see Fig.~\ref{fig1}.
}
\vspace{0.5cm}
\end{figure}

In addition to graphs shown in Fig.~\ref{fig1}, 
diagrams (a)--(c) and (d)--(i) in Fig.~\ref{fig2} formally contribute to the leading ($\nu = 4$) and subleading 
($\nu = 5$) isospin--breaking 3NF, respectively. Their pertinent contributions, however, vanish.  
In particular, for graphs (a), (b), (d), (f) and (i) one observes similar cancellation between various 
time orderings as in the case of the corresponding isospin--invariant 3NFs, see e.g.~\cite{Epelbaum:1998ka}.
Further, the diagram (c) is suppressed by a factor of $q/m$ due to the time derivative entering the Weinberg--Tomozawa
vertex in eq.~(\ref{lagr}). Explicit evaluation of the contributions of graphs (e) and (f) can be performed 
along the lines of ref.~\cite{Epelbaum:2004xf}, which leads to vanishing sum of these contributions.

In \cite{Epelbaum:2004xf} we have  estimated the relative strength of the leading and subleading
corrections compared to the isospin--conserving 3NF at the same order. Isospin--violating 
3NFs are expected to provide a small but non--negligible contribution to the 
$^3$He--$^3$H binding--energy difference.

\section{Summary}
\label{sec:summary}

I have discussed the leading and subleading isospin--violating 3NFs.
The leading contributions are generated by one-- and two--pion exchange diagrams
with their strength given by the strong neutron--proton mass difference. The
subleading corrections are again given by one-- and two--pion exchange diagrams,
driven largely by the charged--to--neutral pion mass difference and also
by the electromagnetic neutron--proton mass difference and the dimension two
electromagnetic LEC $f_1$. In the future, these isospin--breaking 
forces should be used to analyze few--nucleon systems based on chiral EFT.


\begin{theacknowledgments}
I would like to thank Ulf-G.~Mei{\ss}ner for sharing his insight.
This work has been supported by the 
U.S.~Department of Energy Contract No.~DE-AC05-84ER40150 under which the 
Southeastern Universities Research Association (SURA) operates the Thomas Jefferson 
National Accelerator Facility. 
\end{theacknowledgments}


\bibliographystyle{aipproc}   


\begin{thebibliography}{14}
\expandafter\ifx\csname natexlab\endcsname\relax\def\natexlab#1{#1}\fi
\providecommand{\enquote}[1]{``#1''}
\expandafter\ifx\csname url\endcsname\relax
  \def\url#1{\texttt{#1}}\fi
\expandafter\ifx\csname urlprefix\endcsname\relax\def\urlprefix{URL }\fi

\bibitem[van Kolck(1993)]{VanKolck:1993ee}
van Kolck, U.~L., Ph.D. thesis, University of Texas, Austin, USA (1993),
  uMI-94-01021.

\bibitem[van Kolck et~al.(1996)]{vanKolck:1996rm}
van Kolck, U., Friar, J.~L., and Goldman, T., \emph{Phys. Lett.},
  \textbf{B371}, 169--174 (1996).

\bibitem[van Kolck et~al.(1998)]{vanKolck:1997fu}
van Kolck, U., et~al., \emph{Phys. Rev. Lett.}, \textbf{80}, 4386--4389 (1998).

\bibitem[Epelbaum and Mei{\ss}ner(1999)]{Epelbaum:1999zn}
Epelbaum, E., and Mei{\ss}ner, U.-G., \emph{Phys. Lett.}, \textbf{B461},
  287--294 (1999).

\bibitem[Walzl et~al.(2001)]{Walzl:2000cx}
Walzl, M., Mei{\ss}ner, U.~G., and Epelbaum, E., \emph{Nucl. Phys.},
  \textbf{A693}, 663--692 (2001).

\bibitem[Friar and van Kolck(1999)]{Friar:1999zr}
Friar, J.~L., and van Kolck, U., \emph{Phys. Rev.}, \textbf{C60}, 034006
  (1999).

\bibitem[Friar et~al.(2003)]{Friar:2003yv}
Friar, J.~L., van Kolck, U., Payne, G.~L., and Coon, S.~A., \emph{Phys. Rev.},
  \textbf{C68}, 024003 (2003).

\bibitem[Friar et~al.(2004{\natexlab{a}})]{Friar:2004ca}
Friar, J.~L., van Kolck, U., Rentmeester, M. C.~M., and Timmermans, R. G.~E.,
  \emph{Phys. Rev.}, \textbf{C70}, 044001 (2004{\natexlab{a}}).

\bibitem[Epelbaum et~al.(2004)]{Epelbaum:2004xf}
Epelbaum, E., Mei{\ss}ner, U.-G., and Palomar, J.~E., nucl-th/0407037.

\bibitem[Friar et~al.(2004{\natexlab{b}})]{Friar:2004rg}
Friar, J.~L., Payne, G.~L., and van Kolck, U., nucl-th/0408033.

\bibitem[Bernard et~al.(1995)]{Bernard:1995dp}
Bernard, V., Kaiser, N., and Mei{\ss}ner, U.-G., \emph{Int. J. Mod. Phys.},
  \textbf{E4}, 193--346 (1995).

\bibitem[Fettes and Mei{\ss}ner(2001)]{Fettes:2001cr}
Fettes, N., and Mei{\ss}ner, U.-G., \emph{Nucl. Phys.}, \textbf{A693}, 693--709
  (2001).

\bibitem[Meissner and Steininger(1998)]{Meissner:1997ii}
Meissner, U.~G., and Steininger, S., \emph{Phys. Lett.}, \textbf{B419},
  403--411 (1998).

\bibitem[Epelbaum et~al.(1998)]{Epelbaum:1998ka}
Epelbaum, E., Gloeckle, W., and Mei{\ss}ner, U.-G., \emph{Nucl. Phys.},
  \textbf{A637}, 107--134 (1998).

\end{thebibliography}

\end{document}